\documentclass[reqno]{amsart}
\usepackage{amsmath,amssymb,cite}
\usepackage[mathscr]{euscript}
\usepackage{mathtools}
\mathtoolsset{showonlyrefs}
\usepackage{xcolor}
\usepackage{hyperref}

\usepackage[normalem]{ulem}

\definecolor{bblue}{rgb}{.2,0.2,.8}

\theoremstyle{plain}

\theoremstyle{definition}

\theoremstyle{remark}

\numberwithin{equation}{section}
\numberwithin{theorem}{section}

\def\be{\begin{equation}}
\def\ee{\end{equation}}
\def\bp{\begin{pmatrix}}
	\def\ep{\end{pmatrix}}
\def\bea{\begin{eqnarray}}
\def\eea{\end{eqnarray}}

\def\\{\par\medskip}

\let\0=\noindent

\renewcommand{\epsilon}{\varepsilon}
\renewcommand{\hat}{\widehat}

\title{On the equivalence and optimality of transformations of diffusive systems}

\author{Davide Gabrielli}
\address{\noindent Davide Gabrielli \hfill\break\indent
	DISIM, Universit\`a dell'Aquila
	\hfill\break\indent
	67100 Coppito, L'Aquila, Italy
}
\email{davide.gabrielli@univaq.it}

\author{Giovanni Jona-Lasinio}
\address{\noindent Giovanni Jona-Lasinio \hfill\break\indent
	Sapienza Universit\`a di Roma
	\hfill\break\indent
	Diaprtimento di Fisica and INFN
	\hfill\break\indent
	Piazzale Aldo Moro, 5 00185 Roma, Italy
}
\email{gianni.jona@roma1.infn.it}

\date{}
\begin{document}

	\begin{abstract}

		 In this paper we introduce, inspired by Clausius and developing the ideas of \cite{pre}, the concept of equivalence of transformations in non equilibrium theory of diffusive systems within the framework of macroscopic fluctuation theory.   Besides providing a new proof of a formula derived in \cite{mft,qc}, which is the basis of the equivalence, we show that equivalent quasistatic transformations can be distinguished in finite terms, by the renormalized work introduced in \cite{45,46,mft,qc}. This allows us to tackle the problem of determining the optimal quasistatic transformation among the equivalent ones.

	\end{abstract}
\noindent

\maketitle
\thispagestyle{empty}
	\section{Introduction}
	 
	In previous papers \cite{mft,qc} we have shown that for
	diffusive systems one can introduce a notion of renormalized work by
	subtracting the energy necessary to keep the system out of equilibrium.  This interprets in a natural way, within the Macroscopic Fluctuation Theory (MFT), the concept of {\sl house--keeping heat} formulated in \cite{opa}.
	In this way we have for the renormalized work a non--equilibrium
	Clausius type inequality. To a certain extent our approach is complementary to those of  Hatano-Sasa \cite{HS} and Maes-Netocny \cite{MN}.
	
	\medskip
	
	The existence of  entropy in classical equilibrium thermodynamics implies
	that the integral $\int {\frac {\delta Q}{T}}$ does not depend on the
	particular reversible quasi-static transformation joining two fixed
	initial and final states. For example in the simple case of an ideal gas
	we have the exact 1-form
	\begin{equation}
	dS = {\frac {\delta Q}{T}}=C_p {\frac {dT}{T}} - nR{\frac {dp}{p}}\,.
	\end{equation}
	
	\medskip
	
	In his ninth memoir Clausius summarizes the two principles of
	thermodynamics as follows
	\begin{quotation}
		{\sl The whole mechanical theory of heat rests on two fundamental
			theorems, - that of equivalence of heat and work, and that of equivalence
			of transformations.}\end{quotation}
	 In particular all reversible transformations between
	fixed equilibrium states are equivalent in the sense of Clausius. In processes with
	dissipation it is important to have an optimality criterium. This was introduced in \cite{qc} in terms of the renormalized work. Practically the type of transformation in a process with dissipation may be dictated by experimental possibilities, the problem is then to find among all equivalent transformations, the best one. A solvable example was discussed in \cite{qc}.
	
	\smallskip
	
	Previously \cite{pre} we pointed out that in the Macroscopic Fluctuation
	Theory of diffusive systems a formula containing only finite quantities
	obtained in \cite{mft,qc} implies the existence of an exact 1-form also in
	a dissipative context. This in turn implies an equivalence relationship of
	quasi--static transformations, possibly very different energetically:
	this point was not discussed in our previous papers and here we extend the discussion in \cite{pre}.
	The equivalence in the sense of Clausius \cite{c4} can be
	extended to stationary non equilibrium states of diffusive systems.
	
	\medskip
	
	It may be surprising an equivalence relation in dissipative irreversible
	transformations which may differ in a quasi static limit by an infinite
	amount of dissipated energy. As we will show in the following two
	equivalent transformations can be distinguished in finite terms by the
	renormalized work which depends, in an asymptotic region in time
	\cite{qc}, on the protocols of the transformations.
	
	\medskip
	
	In the present paper, after recalling some basic facts of the MFT, we
	provide a different proof of the non--equilibrium equivalence relation and
	discuss a result on optimality of a transformation obtained in \cite{qc}.
	We then illustrate various aspects of this non--equilibrium equivalence
	with the help of examples. 

\medskip

Various definitions of renormalized work have been introduced for example by \cite{HS,MN}. What is of interest for us in the present context is the case of MFT. A comparison between \cite{HS} and \cite{MN} is made in this last work. The difference between our definition of the the renormalized work and that of \cite{MN} has been discussed in our paper \cite{46}. We emphasize later that a solvable example discussed in \cite{qc} is a case of selecting the optimal transformation between two equivalent ones.  Of course in concrete cases it may be difficult to solve the related variational principle analytically and we have to resort to numerical schemes.

\section{ Macroscopic Dynamics}
 We denote by $\Lambda \subset \mathbb R^d$ the
bounded region occupied by the system, by $\partial \Lambda$ the
boundary of $\Lambda$, by $x$ the macroscopic space coordinates and by
$t$ the macroscopic time. The system is in contact with boundary
reservoirs, characterized by their chemical potential $\lambda (t,x)$, $x \in \partial \Lambda$ and under the action of an external field $E(t,x)$, $x\in \Lambda$.

\medskip

At the macroscopic level the assumption is that the system is completely described by the local density $\rho(t,x)$ and the local density current $j(t,x)$ and their evolution is given by the continuity equation together with the constitutive equation which expresses the current as a function of the density. Namely,
\begin{equation}
\label{2.1}
\begin{cases}
\partial_t \rho (t) + \nabla\cdot j (t) = 0,\\
j (t)= J(t,\rho(t)),
\end{cases}
\end{equation}
where we omit the explicit dependence on the space variable $x\in\Lambda$.
For driven diffusive systems the constitutive equation takes the form
\begin{equation}
\label{2.2}
J(t,\rho)  = - D(\rho) \nabla\rho + \chi(\rho) \, E(t),
\end{equation}
where the \emph{diffusion coefficient} $D(\rho)$ and the \emph{mobility}
$\chi(\rho)$ are assumed to be $d\times d$ symmetric and positive definite matrices. This holds in the context of stochastic lattice gases \cite{mft}. We will consider in this paper just the cases where $D$ and $\chi$ are multiple of the identity matrix, so that $D,\chi$ can be considered as scalars. Equation \eqref{2.2} relies  on the local equilibrium hypothesis, small local gradients and linear response to the external field. We also assume that the time dependence of the external fields and chemical potentials is on a scale which is long with respect to the relaxation time. Otherwise we may expect that the hydrodynamic equations do not describe correctly the physics of the system.

The evolution of the density is thus given  by the driven diffusive equation
\begin{equation}
\label{r01}
\partial_t \rho (t) + \nabla\cdot \big( \chi(\rho)  E(t) \big)
= \nabla\cdot \big( D(\rho) \nabla\rho \big).
\end{equation}
The transport coefficients $D$ and $\chi$ satisfy the local Einstein relation
\begin{equation}
\label{r29}
D(\rho) = \chi(\rho) \, f''(\rho),
\end{equation}
where $f$ is the equilibrium free energy per unit  volume which we assume
to depend on the local value  $\rho (x)$.

\medskip

Equations \eqref{2.1}--\eqref{2.2} have to be supplemented by the
appropriate boundary conditions on $\partial\Lambda$ due to the
interaction with the external reservoirs. If $\lambda(t,x)$,
$x\in\partial \Lambda$, is the chemical potential of the external
reservoirs, the boundary condition reads
\begin{equation}
\label{2.3}
f'\big(\rho(t,x) \big) = \lambda(t,x), \qquad x\in\partial \Lambda.
\end{equation}

\smallskip
We observe that given any density profile $\rho(x,t)$ and any current $j(x,t)$, then there exists an external field $E(x,t)$ such that $j=J(\rho)$, this is given indeed by
\begin{equation}\label{campo-E}
E(t)=\frac{j(t)-D(\rho(t))\nabla \rho(t)}{\chi(\rho(t))}\,.
\end{equation}
This fact implies that any pair $(\rho(x,t),j(x,t))$ related by the continuity equation
can be realized as possible space time trajectory of the system for suitable external fields.

\subsection{Equilibrium and non equilibrium states}

We now introduce some terminology for non--equilibrium states.

\subsubsection{Dynamical states}
We call a \emph{dynamical state of the system}, in the time window $[0,T]$, any pair $(\rho(t),j(t))_{t\in[0,T]}$, connected by the continuity equation \eqref{2.1}, so that in the non equilibrium situation the state of the system is rather a trajectory.

\medskip

There is a natural one to one correspondence between the pair $(\rho(t),j(t))_{t\in [0,T]}$ and the triple $\Big((\lambda(t),E(t))_{t\in[0,T]},\rho(0)\Big)$ where the third element is the initial condition for the density.

Given the pair $(\rho(t),j(t))_{t\in [0,T]}$, the initial condition $\rho(0)$ is given and we obtain the pair $(\lambda(t),E(t))_{t\in[0,T]}$ by the equations \eqref{2.3} \eqref{campo-E}. Conversely given $(\lambda(t),E(t))_{t\in[0.T]}$ and $\rho(0)$, we obtain $(\rho(t),j(t))_{t\in[0,T]}$ solving the hydrodynamic equation that is summarized as
\begin{equation}
\left\{
\begin{array}{ll}
\partial_t \rho (t) = \nabla\cdot \big( D(\rho) \nabla\rho \big)- \nabla\cdot \big( \chi(\rho)  E(t) \big)\,, & x\in \Lambda\,,\\
f'\big(\rho(t,x) \big) = \lambda(t,x)\,, & x\in\partial \Lambda\,,\\
j(x,t)=- D(\rho(x,t)) \nabla\rho(x,t) + \chi(\rho(x,t)) \, E(x,t)\,, & x\in \Lambda\,,
\end{array}
\right.
\end{equation}
adding the initial condition $\rho(0)$.

\subsubsection{Stationary states}

If the chemical potential and external field do not depend on time, we
denote by $\bar\rho=\bar\rho_{\lambda, E}$ the stationary solution of
\eqref{r01},\eqref{2.3}. The stationary density profile $\bar\rho$ is characterized by the vanishing of the divergence of the associated current, $\nabla \cdot J(\bar\rho)=0$
\begin{equation}
\label{05}
\begin{cases}
\nabla \cdot J(\bar\rho)= \nabla \cdot \big( -D(\bar\rho)
\nabla\bar\rho + \chi(\bar\rho) \, E  \big) = 0,
\\
 f' (\bar\rho(x)) = \lambda (x),
\qquad x\in\partial \Lambda.
\end{cases}
\end{equation}
We will assume that the stationary solution is unique. The stationary states furthermore split into \emph{equilibrium stationary states} and \emph{non equilibrium stationary states}. As we will explain, the equilibrium states are characterized by the condition $\bar J=J(\bar\rho)=0$. The following arguments are a reformulation of those in \cite{tt}.

\smallskip

We examine the correspondence between variables $(\lambda,E)$ and the stationary solution $(\bar\rho,\bar J)=(\bar\rho_{\lambda,E}, \bar J_{\lambda, E})$. As in the dynamical case, we have a one to one correspondence, that we now examine in some detail. Given a pair $(\lambda,E)$ the (assumed) unique corresponding pair $(\bar\rho, \bar J)$ is obtained solving \eqref{05}. Conversely, given a pair $(\bar\rho, \bar J)$ such that $\nabla\cdot \bar J=0$, the values of the chemical potential at the boundary are obtained by the second relation in \eqref{05}, while the external field $E$
is obtained by a stationary version of equation \eqref{campo-E}
\begin{equation}\label{campo-E-staz}
E=\frac{\bar J-D(\bar\rho)\nabla \bar\rho}{\chi(\bar\rho)}\,.
\end{equation}

In particular if we fix only the density stationary profile $\bar\rho$ we may have several corresponding variables $(\lambda, E)$. A special situation is when the current itself vanishes, $J(\bar\rho)=0$; if this is the case, which means that the boundary conditions and the external field balance each other, we say that the system is in a macroscopic equilibrium state; this can be viewed as a macroscopic counterpart to detailed balance \cite{tt}.

\medskip

We may have both homogeneous and non homogeneous equilibrium states. The homogeneous equilibrium states correspond to the case in which the external field vanishes and the chemical potential is constant in space, in this case also the stationary solution $\bar\rho$ is constant in space. We may have more general equilibrium states that are the inhomogeneous equilibrium states and correspond to the case in which the external field is gradient, $E=\nabla U$, and it is possible to choose the arbitrary constant in the definition of $U$ such that $U(x)=\lambda(x)$, $x\in\partial\Lambda$. We emphasize that if this second condition is not satisfied, the state is non equilibrium and the current $\bar J$ is not zero, even if the external field is of gradient type. This follows by the following argument.

\smallskip

We have an equilibrium state, in any dimension, when the stationary condition $\nabla\cdot J(\bar\rho)=0$ is satisfied with $J(\bar\rho)=0$. Using the form of the current given by the constitutive equation \eqref{2.2}, this corresponds to the condition
\begin{equation}
-D(\bar\rho)\nabla \bar\rho+\chi(\bar\rho)E=0\,,
\end{equation}
that by the local Einstein relation becomes
\begin{align}
E=\chi^{-1}(\bar\rho) D(\bar\rho)\nabla \bar\rho=f''(\bar\rho)\nabla\bar\rho=\nabla f'(\bar\rho)\,.
\end{align}
This means that for equilibrium states the external field $E$ has to be of gradient type $E=\nabla U$ and in particular we have that $U(x)=f'(\bar\rho(x))+c$ for a constant $c$. By the boundary conditions that have to be satisfied in all cases, $f'(\bar\rho(x))=\lambda(x)$ $x\in \partial \Lambda$, we have that $U(x)=\lambda(x)+c$ when $x\in \partial \Lambda$ and $c$ is a suitable constant.

Conversely if we have $E(x)=\nabla U(x)$, the equation that characterizes the equilibrium states $J(\bar\rho)=0$ becomes
\begin{equation}\label{stzU}
-D(\bar\rho(x))\nabla \bar \rho(x)+\chi(\bar\rho(x))E(x)=0\,, \qquad x\in \Lambda\,,
\end{equation}
that, by the Einstein relation is equivalent to $\nabla U(x)=\nabla f'(\bar\rho(x))$ whose general solution is $f'(\bar\rho(x))=U(x)+K$ for an arbitrary constant $K$. Imposing the boundary condition we obtain $\lambda(x)=U(x)+K$, $x\in \partial \Lambda$. When this is satisfied, for a given constant $K^*$, then the stationary density is given by $\bar\rho(x)=(f')^{-1}(U(x)+K^*)$. If instead there is not such a constant $K^*$ then it is not possible to find a $\bar\rho$ satisfying \eqref{stzU} and the boundary conditions. This implies that the stationary current $\bar J\neq 0$ and the stationary state is a non equilibrium one.

\smallskip

If we fix the stationary density profile $\bar\rho$ and search for all the possible external fields that produce such a stationary profile, this class can be written as
\begin{equation}\label{E-gen}
E(x)= \nabla f'(\bar\rho(x))+\chi^{-1}(\bar\rho(x))G(x)=E^*(x)+\chi^{-1}(\bar\rho(x))G(x)\,,
\end{equation}
where $E^*(x)$ is defined by the last equality and $G$ is an arbitrary divergence free field.
Among all such external fields the only one for which $(\lambda, E)$ is an equilibrium state is the one that corresponds to $G=0$.

The stationary equation $\nabla \cdot J = 0$ with the associated boundary conditions play a role akin to the equation of state in equilibrium thermodynamics.

\medskip

\section{Transformations and Energy Balance}

Consider a system in a time dependent environment, that is, $E$ and $\lambda$ depend on time. The work done by the environment on the system in the time interval $[0,T]$ is
\begin{equation}
  \label{W=}
  \begin{split}
    W_{[0,T]} = & \int_{0}^{T}\! dt \, \Big\{ \int_\Lambda \!dx\, j(t)
    \cdot E(t)
    - \int_{\partial\Lambda} \!d\sigma \, \lambda
    (t) \: j(t) \cdot \hat{n} \Big\},
  \end{split}
\end{equation}
where $\hat n$ is the outer normal to $\partial \Lambda$ and $d\sigma$
is the surface measure on $\partial \Lambda$.  The first term on the
right hand side is the energy provided by the external field
while the second is the energy provided by the reservoirs.

\medskip

Fix time dependent paths $\lambda(t)$ of the chemical potential and
$E(t)$ of the field. Given a density profile $\rho_0$, let
$\rho(t)$, $j(t)$, $t \ge 0$, be the solution of
\eqref{2.1}--\eqref{2.3} with initial condition $\rho_0$.
By using the Einstein relation \eqref{r29} and the boundary condition
$f'(\rho(t)) = \lambda(t)$, an application of the
divergence theorem yields
\begin{equation}
\label{04}
W_{[0,T]} \,=\,  F(\rho(T)) - F(\rho(0))
+\,  \int_{0}^{T} \!dt  \int_\Lambda \!dx \;
j(t)\cdot \chi(\rho(t) )^{-1} j(t),
\end{equation}
where $F$ is the free energy functional of equilibrium thermodynamics,
\begin{equation}
\label{10}
F(\rho) = \int_\Lambda \!dx \: f (\rho(x)).
\end{equation}
We emphasize that the functional $F(\rho)$ can be evaluated on any profile $\rho(x)$ and does not depend on the current. In particular can be evaluated on the stationary states $\bar\rho$ and the result is independent of the stationary current $\bar J$.
The step from \eqref{W=} to \eqref{04} uses  the constitutive equation  \eqref{2.2} and the Einstein relationship \eqref{r29}. It tells us that for the class of systems considered the change of the free energy $\Delta F$ is, as expected, the difference between the total work and the total dissipation. Notice that if the system is out of equilibrium for an infinite time both these quantities are infinite. The currents $j(t)$ must be evaluated on the solutions of the hydrodynamic equations.

\medskip

From \eqref{04} follows the Clausius inequality
\begin{equation}
W_{[0,T]} \,\geq\,  F(\rho(T)) - F(\rho(0))
\end{equation}
For quasi-static transformations between nonequilibrium stationary states, the Clausius inequality  does not carry any significant information as discussed for example in \cite{qc}.

\subsection{Quasistatic transformations}

To analyze transformations over a long interval $[0,\tau]$ driven by slowly changing boundary conditions and external fields, it is convenient to rescale time and introduce the rescaled variable $s={t}/{\tau}$. The \emph{protocol} of a transformation  is defined therefore by a choice of the external drivings $E(s,x)$, $x\in\Lambda$, and $\lambda(s,x)$, $x\in \partial\Lambda$, $s\in[0,1]$. The driving terms are assumed to be $C^1$ in the parameter $s\in [0,1]$ and to satisfy the conditions
\begin{equation}\label{derEl=0}
\left\{
\begin{array}{l}
\dot E(0)=\dot E(1)=0\,,\\
\dot\lambda(0)=\dot\lambda(1)=0\,.
\end{array}
\right.
\end{equation}
This is a natural condition imagining that the drivings before and after the transformation are time independent.

\smallskip

A real transformation that takes place in the time window $[0,\tau]$ is the solution $\left(\rho^\tau(x,t),j^\tau(x,t)\right)_{x\in \Lambda, t\in[0,\tau]}$ of the hydrodynamic equation
\begin{equation}
\left\{
\begin{array}{ll}
\partial_t \rho^\tau (t) = \nabla\cdot \big( D(\rho^\tau) \nabla\rho^\tau \big)- \nabla\cdot \big( \chi(\rho^\tau)  E^\tau(t) \big) & x\in \Lambda\,, \ t\in[0,\tau]\\
f'\big(\rho^\tau(t,x) \big) = \lambda^\tau(t,x), & x\in\partial \Lambda\,, \ t\in [0,\tau]\\
j^\tau(x,t)=- D(\rho^\tau(x,t)) \nabla\rho^\tau(x,t) + \chi(\rho^\tau(x,t)) \, E^\tau(x,t) & x\in \Lambda\,,\ t\in [0,\tau]\,,
\end{array}
\right.
\end{equation}
where the external drivings $\left(E^\tau(x,t), \lambda^\tau(x,t)\right)_{x\in \Lambda, t\in[0,\tau]}$ are defined by $E^\tau(x,t):=E(x,t/\tau)$, $\lambda^\tau(x,t):=\lambda(x,t/\tau)$.

\smallskip

Our analysis is based in the following asymptotic expansion for large $\tau$
\begin{equation}\label{davide}
\begin{array}{l}
\displaystyle{
\rho^\tau (\tau s)=\bar \rho(s)+\tfrac 1\tau \, r(s)
+ o\big(\tfrac1{\tau}\big)\,,
} \\
\displaystyle{
j^\tau(\tau s)
=J(s,\bar\rho(s))+\tfrac 1\tau \, g(s) + o\big(\tfrac 1{\tau} \big)
\,,}\\
\end{array}
\end{equation}
that is physically reasonable and in some cases can be verified to hold. .

It is easy to see that $r,g$ obey the following equations
\begin{equation}\label{davide2bis}
\left\{
\begin{array}{l}
\partial_s  \bar \rho(s) + \nabla \cdot g(s) =0 \\
g(s)= -
\nabla \Big( D(\bar\rho(s)) r(s) \Big)
+ r(s) \chi'(\bar \rho(s)) E(s)
\\
r(s, x)=0, \; x\in\partial \Lambda\,\\
r(x,0)=r(x,1)=0\,, \; x\in \Lambda\,,
\end{array}
\right.
\end{equation}
which implies that $r,g$ depend linearly on $\partial_s \bar\rho$ and are actually proportional to it. The equations on the last line follows by the conditions \eqref{derEl=0}

\medskip

We now consider the identity \eqref{04}, using also the definition \eqref{W=}, for the transformation
$\left(\rho^\tau(x,t),j^\tau(x,t)\right)_{x\in \Lambda, t\in[0,\tau]}$ and perform an expansion on the small parameter $1/\tau$ on both sides. We have equality among all the coefficients of the expansions on the two sides. The first non trivial relation, that corresponds to the zero order in $1/\tau$ is given by
\begin{equation}
\label{f}
  \begin{split}
F\big(\bar \rho(1)\big)-F\big(\bar \rho(0)\big)
    &\; = \int_0^1\!ds \int_\Lambda\!dx \, E(s)\cdot g(s) -\int_0^1\!ds
    \int_{\partial \Lambda}\! d\sigma \,\lambda(s) g(s)\cdot \hat n
    \\
    &\;\; + \int_0^1\!ds \int_\Lambda \!dx\, r(s)
    J(s,\bar\rho(s))\cdot (\chi^{-1})'\big(\bar \rho(s)\big)
    J(s,\bar\rho(s))
    \,.
  \end{split}
\end{equation}
Equation \eqref{f} establishes a mathematical equivalence of quasi--static transformations for a class of dissipative systems. These transformations can be energetically  very different as the energy necessary to keep the system out of equilibrium per unit of time can differ considerably. This relation gives the variation of equilibrium free energy in terms of an integration on quasistatic transformations with suitable variations $(r,g)$ of density and current, the result is independent of the specific transformation. The first two terms of this formula represent a linearization of the total work along the transformation, while the third one is due to the variation of the mobility which in general will be a nonlinear function of the density. There is no corresponding term for the diffusion coefficient $D$ since on the scale $\tau$ the relaxation of the system is instantaneous \cite{qc}.

For a detailed derivation using a slow limit and an expansion we refer to \cite{mft,qc}. Here we give a direct proof.

\smallskip

\begin{proof}[Proof of equation (3.5)]
The left hand side of \eqref{f} can be written as $\int_0^1\!ds\int_\Lambda\!dx \frac{\delta F}{\delta \bar\rho(s)}\partial_s \bar\rho(s)=\int_0^1\!ds\int_\Lambda\!dx f'(\bar\rho(s))\partial_s\bar\rho(s)$. To prove equality \eqref{f} it is enough therefore to verify
\begin{align*}
\int_\Lambda\!dx f'(\bar\rho(s))\partial_s\bar\rho(s)&= \int_\Lambda\!dx \, E(s)\cdot g(s) -
\int_{\partial \Lambda}\! d\sigma \,\lambda(s) g(s)\cdot \hat n
\\
&\;\; + \int_\Lambda \!dx\, r(s)
J(s,\bar\rho(s))\cdot (\chi^{-1})'\big(\bar \rho(s)\big)
J(s,\bar\rho(s))\,.
\end{align*}
Using the first equation in \eqref{davide2bis} an integration by parts and the boundary conditions of the hydrodynamics we get that the left hand side of the above equation is given by
$$
-
\int_{\partial \Lambda}\! d\sigma\,\lambda(s) g(s)\cdot \hat n +\int_\Lambda\!dx f''(\bar\rho(s))g\cdot \nabla \bar\rho(s)\,.
$$
It remains therefore to show
$$
\int_\Lambda\!dx \left[f''(\bar\rho(s))\nabla\bar\rho(s)-E\right]\cdot g=-\int_\Lambda \!dx\, r(s)
J(s,\bar\rho(s))\cdot \frac{\chi'(\bar\rho(s))}{\chi^2(\bar\rho(s))}
J(s,\bar\rho(s))\,.
$$
We can now elaborate the left hand side as follows
\begin{align*}
&\int_\Lambda\!dx \left[f''(\bar\rho(s))\nabla\bar\rho(s)-E\right]\cdot g
= -\int_\Lambda\!dx \frac{J(\bar\rho(s))}{\chi(\bar\rho(s))}\cdot g \\
= &\int_\Lambda\!dx \frac{J(\bar\rho(s))}{\chi(\bar\rho(s))}\cdot \Big(\nabla(D(\bar\rho(s))r(s))-r(s)\chi'(\bar\rho(s))E(s)\Big)\\
=& \int_\Lambda\!dx \Big[D(\bar\rho(s))r(s)\frac{\chi'(\bar\rho(s))\nabla\bar\rho(s)}{\chi^2(\bar\rho(s))}\cdot J(\bar\rho(s))-r(s)\frac{\chi'(\bar\rho(s))E(s)}{\chi(\bar\rho(s))}\cdot J(\bar\rho(s))\Big]\\
=&-\int_\Lambda \!dx\, r(s)
J(s,\bar\rho(s))\cdot \frac{\chi'(\bar\rho(s))}{\chi^2(\bar\rho(s))}
J(s,\bar\rho(s))\,.
\end{align*}
In the first equality we used the Einstein relation, in the second equality the definition of $g$, in the third equality we did an integration by parts, used the fact that $r|_{\partial\Lambda}=0$ and the fact that $\nabla\cdot J(\bar\rho(s))=0$, and finally in the last equality we used just the definition of $J(\bar\rho(s))$. This finishes the proof.
\end{proof}

\smallskip
We observe  that equations \eqref{davide2bis} are invariant under the following time reversal transformations
\begin{equation}
\left\{
\begin{array}{l}
\bar\rho(s)\to \bar\rho(1-s)\\
r(s)\to -r(1-s)\\
g(s)\to -g(1-s)
\end{array}
\right.
\end{equation}
and moreover the right hand side of formula \eqref{f} changes sign under the above transformations. This is an indication of the reversibility of quasistatic transformations.

\subsection{States and free energy}

A transformation corresponds to a path $\gamma$ in the infinite dimensional space of the variables $(E,\lambda)$ that completely identify the state. We recall that since $\bar\rho$ depends of the external drivings $(E,\lambda)$, we have that
$J(\bar\rho)$ and $F(\bar\rho)$ are functions of the variables $(E,\lambda)$.
With a small abuse of notation we call again
$F(\lambda,E):=F(\bar\rho_{E,\lambda})$, the current and free energy written in terms of the external drivings $(\lambda,E)$.

Given a curve $\gamma$ on the space of states such that $(\lambda(0),E(0))=\gamma(0)$ and $(\lambda(1),E(1))=\gamma(1)$ we obtain a line integral of an exact 1--form
\begin{equation}
 \label{oqc}
 F(E(1),\lambda(1))- F(E(0),\lambda(0)) = \int_{\gamma}   \frac{\delta  F}{\delta E} d E+ \frac{\delta  F}{\delta \lambda} d\lambda\,.
\end{equation}
This formula corresponds to an integral version of the differential of the free energy $ F$ written as
\begin{equation}\label{diff}
dF=\int_\Lambda dx \frac{\delta  F}{\delta E}(x)\delta E(x)
+\int_{\partial \Lambda}d\sigma(x)\frac{\delta  F}{\delta \lambda}(x)\delta\lambda(x) \,.
\end{equation}

\subsection{The renormalized work}
\label{sec:ren-w}

\subsubsection{Decomposition of the current}
In the framework of macroscopic fluctuation theory
it is possible to define a thermodynamic functional $V$,
called the quasi-potential, and coinciding with a large deviations rate functional of the stationary measure.
It can be characterized
as the maximal positive solution, vanishing when $\rho=\bar \rho$,
of the infinite dimensional Hamilton-Jacobi equation,
\begin{equation}
\label{r15-2}
\int_{\Lambda}dx\,  \nabla \frac{\delta V}{\delta \rho} \cdot \chi(\rho)
\nabla \frac{\delta V}{\delta \rho}
-  \int_{\Lambda}dx\, \frac{\delta V}{\delta \rho} \nabla \cdot J(\rho)
= 0\,.
\end{equation}

Using $V$ it is possible to define the \emph{symmetric current} $J_S$ by
\begin{equation}
\label{Ons-form}
J_S(\rho)=-\chi(\rho)\nabla\frac{\delta V}{\delta\rho}.
\end{equation}
Since the stationary density $\bar\rho$ is a minimum for $V$, then
$(\delta V/\delta\rho)(\bar\rho)=0$.  The symmetric current thus
vanishes at the stationary profile,
\begin{equation}
\label{01}
J_S(\bar\rho) \;=\;0.
\end{equation}
We rewrite the hydrodynamic current as
\begin{equation}
\label{split}
J(\rho)=J_S(\rho)+J_A(\rho),
\end{equation}
which defines the \emph{antisymmetric current} $J_A$.

In view of these definitions, the Hamilton Jacobi equation \eqref{r15-2} becomes
\begin{equation}
\label{r18}
\int_{\Lambda}dx\,  J_S (\rho) \cdot \chi(\rho)^{-1} J_A(\rho)
= 0\,,
\end{equation}
that is an orthogonality condition with respect to the metric $\chi^{-1}$.

In the case of an equilibrium state the quasi-potential
$V=V_{\lambda,E}(\rho)$ is the local functional
\begin{equation}\label{localV}
V_{\lambda,E}(\rho)
=
\int_\Lambda dx
\big(
f(\rho)-f(\bar\rho)-f^\prime(\bar\rho)(\rho-\bar\rho)
\big)
\,,
\end{equation}
where $\bar\rho=\bar\rho_{\lambda,E}$ is the  solution of \eqref{05}.

\subsubsection{Renormalized work}

In the case of a non equilibrium stationary state,
we define a renormalized work subtracting
the energy needed to maintain the system out of equilibrium.
For time independent drivings, by
the orthogonal decomposition \eqref{split} and \eqref{01},
$J(\bar\rho)=J_{\mathrm{A}} (\bar\rho)$ is the macroscopic current in
the stationary state.  In view of the general formula for the total
work \eqref{04}, the amount of energy per unit time needed to maintain
the system in the stationary profile $\bar\rho$ is
\begin{equation}
\label{toman}
\int_\Lambda \!dx \:
J_\mathrm{A}(\bar\rho) \cdot \chi(\bar\rho)^{-1}
J_\mathrm{A}(\bar\rho).
\end{equation}

Fix now $T>0$, a density profile $\rho_0$, and space-time dependent
chemical potentials $\lambda(t)$ and external field $E(t)$, $t\in
[0,T]$.  Let $(\rho(t),j(t))$ be the corresponding solution of
\eqref{2.1}--\eqref{2.3} with initial condition $\rho_0$.
We define the renormalized work
$W^\textrm{ren}_{[0,T]}$ done by the reservoirs and the external field
in the time interval $[0,T]$ as
\begin{equation}
\label{Weff}
W^\textrm{ren}_{[0,T]} = W_{[0,T]}
- \int_{0}^{T}\! dt \int_\Lambda \!dx\, J_\mathrm{A}(t,\rho(t)) \,
\cdot \chi(\rho(t))^{-1} J_\mathrm{A}(t,\rho(t))
\end{equation}
where $J_\mathrm{A}(t,\rho)$ is the antisymmetric current for the
system with the time independent external driving obtained by freezing
the time dependent chemical potential $\lambda$ and external field $E$
at time $t$.  Observe that the definition of the renormalized
work involves the antisymmetric current $J_\mathrm{A}(t)$ computed not
at density profile $\bar\rho_{\lambda(t), E(t)}$ but at the solution
$\rho(t)$ of the time dependent hydrodynamic equation.

The definition \eqref{Weff} is natural within the macroscopic
fluctuation theory and leads to a Clausius inequality.  Indeed, in
view of \eqref{04} and the orthogonality in \eqref{r18} between the
symmetric and the antisymmetric part of the current,
\begin{equation}
\label{Weff2}
\begin{split}
W^\textrm{ren}_{[0,T]} & = F(\rho(T)) - F(\rho_0)
+ \int_{0}^{T}\! dt \int_\Lambda \!dx\, J_\mathrm{S}(t,\rho(t)) \cdot
\chi(\rho(t))^{-1} J_\mathrm{S}(t,\rho(t))\\
& \geq F(\rho(T)) - F(\rho_0)\,.
\end{split}
\end{equation}

In the limit of slow transformations, the expansion in $1/\tau$ of the renormalized work is \cite{mft,qc}
\begin{equation}
\label{1taub}
W^{\mathrm{ren}}_{[0,\tau]} =
\big[ F(\rho^\tau(\tau))-F(\rho^\tau(0))\big]
+
\frac 1\tau \,B   + O \big(\tfrac 1{\tau^2}\big)
\,,
\end{equation}
where the \emph{excess} functional $B$ is:
\begin{equation}
\label{1tau}
B = \int_0^1\!ds\!\int_\Lambda\!dx \,
\nabla\big(  C_s^{-1} r(s) \big) \cdot
\chi(\bar\rho(s))   \nabla\big(  C_s^{-1} r(s) \big).
\end{equation}
where
$C^{-1}_s(x,y)$ is the inverse of the covariance in the stationary state associated to the external drivings $(\lambda(s),E(s))$.

\medskip

By developing further for large $\tau$
\begin{equation}
F(\bar\rho(1)) - F(\bar\rho(0)) + {\frac 1 \tau} [\partial_{\bar\rho(1)} F(\bar\rho(1)) r(\bar\rho(1)) - \partial_{\bar\rho(0)} F(\bar\rho(0)) r(\bar\rho(0))] + \frac 1\tau \,B
\end{equation}
taking into account that $r$ depends linearly on $\partial_s \bar\rho$ and that the initial and final states are stationary, by \eqref{derEl=0}, only the $\frac 1\tau \,B$ survives which depends on the protocols and allows to distinguish equivalent transformations.

\smallskip

In \cite{mft,qc} we considered two cases of optimal transformations, one through homogeneous equilibrium states and the other through inhomogeneous equilibrium states. These transformations are clearly equivalent according to the definition of the present paper and are sharply distinguished by the value of $B$.

\medskip

In \cite{46} we have shown that in the limit of a quasi--static transformation we have  $W^\mathrm{ren} = \Delta F$ where $\Delta F$ is the difference between  the equilibrium free energies evaluated in the final and initial states of the transformation. From \eqref{oqc} we obtain for large $\tau$ the representations of the renormalized work
\begin{equation}
  W^{ren} = \int_{\gamma}   \frac{\delta  F}{\delta E} d E+ \frac{\delta F}{\delta \lambda} d\lambda\,.
\end{equation}

\section{Examples}
\label{sec:ex}

To illustrate the different terms appearing  in the above formula \eqref{f} we now concentrate on one dimensional examples with constant external field where everything can be made more explicit.
Consider a one dimensional interval $\Lambda=[0,L]$. In this case we have that $F$ is a functional depending on the two boundary values of the chemical potential $\lambda_0$ and $\lambda_L$ and of the constant value of the external field $E$; there is also a dependence on the size $L$ of the system that we do not write explicitly. Without loss of generality we assume $\lambda_0\leq \lambda_L$. In formula \eqref{diff} we have therefore to compute the derivatives $\frac{\partial F}{\partial E}$, $\frac{\partial F}{\partial \lambda_0}$
and $\frac{\partial  F}{\partial \lambda_L}$ and formula \eqref{diff} reads simply as
\begin{equation}\label{diffF}
d F=\frac{\partial F}{\partial E}dE+\frac{\partial  F}{\partial \lambda_0}d\lambda_0+
\frac{\partial F}{\partial \lambda_L}d\lambda_L\,.
\end{equation}

The stationary condition \eqref{05} can be written as
\begin{equation}\label{staz-1d}
\left\{
\begin{array}{ll}
D(\bar\rho)\partial_x\bar\rho= E\chi(\bar\rho)-c\,,\\
f'(\bar\rho(0))=\lambda_0\,, f'(\bar\rho(L))=\lambda_L
\end{array}
\right.
\end{equation}
where $c$ is a constant to be determined by the boundary conditions. Calling $\rho_0:=(f')^{-1}(\lambda_0)$
and $\rho_L:=(f')^{-1}(\lambda_L)$ the constant $c$ is determined by integrating both sides of the first equation in \eqref{staz-1d} obtaining
\begin{equation}\label{condizione}
\int_{\rho_0}^{\rho_L}\frac{D(\rho)d\rho}{E\chi(\rho)-c}=L\,.
\end{equation}
The constant $c$ has the natural interpretation as the current observed in the stationary state, that by the constitutive equation \eqref{2.2} is $J(\bar\rho(x))=c$, which we emphasize does not depend on $x$.
If we define
\begin{equation}
G=G_{c,E,\rho_0}(\rho):=\int_{\rho_0}^{\rho}\frac{D(a)da}{E\chi(a)-c}\,d\,a\,,
\end{equation}
condition \eqref{condizione} reads $G(\rho_L)=L$.
Integrating by separation of variables the stationary condition \eqref{staz-1d}, we get
\begin{equation}\label{staz-sol}
\int_{\rho_0}^{\bar\rho(x)}\frac{D(\rho)d\rho}{E\chi(\rho)-c}=x\,
\end{equation}
that is equivalent to $\bar\rho(x)=G^{-1}(x)$ and we deduce
\begin{equation}\label{freeFgen}
F=\int_0^Lf\Big(G^{-1}_{c,E,\rho_0}(x)\Big)dx\,,
\end{equation}
where we recall that $c=c(E,\lambda)$ is determined by \eqref{condizione}. This is a formula that can give explicitly the dependence of $F$ on the different parameters.

\subsection{Equilibrium states}
\label{eqstate-ex}
Formula \eqref{freeFgen} can be made more explicit in the case of equilibrium states.

\smallskip

We consider first the homogeneous equilibrium states that correspond to the situation $\lambda_0=\lambda_L=\lambda$ and $E=0$. In this case the stationary profile is constant and given by
$\bar\rho=\left(f'\right)^{-1}(\lambda)$. It is natural to define $\hat f:= f \circ \left(f'\right)^{-1}$ that is the density of free energy written in term of the chemical potential. The equilibrium free energy in the homogeneous case is therefore
\begin{equation}
F_{eq}(\lambda,\lambda):=F(0,\lambda,\lambda)=\int_0^L\hat f(\lambda)dx=L\hat f(\lambda)\,.
\end{equation}

\subsubsection{Inhomogeneous equilibrium states}

In the non homogeneous case, the equilibrium situation corresponds to the value $c=0$ in \eqref{condizione}. Using the
Einstein relation in this case such condition becomes
$$
\frac 1E \int_{\rho_0}^{\rho_L}f'(\rho)d\rho=L\,
$$
that fixes the value of the external field to $E_{eq}=E_{eq}(\lambda_0,\lambda_L):=\frac{\lambda_L-\lambda_0}{L}$. If we define $\bar\lambda(x)=f'(\bar\rho(x))$ we obtain by \eqref{staz-sol} $\bar\lambda(x)=\lambda_0+\frac xL(\lambda_L-\lambda_0)$. By a change of variable we obtain that the equilibrium free energy is given by
\begin{equation}\label{questa}
 F_{eq}(\lambda_0,\lambda_L):=F(E_{eq},\lambda_0,\lambda_L)=\frac{L}{\lambda_L-\lambda_0}\int_{\lambda_0}^{\lambda_L} \hat f(\lambda)d\lambda=\frac{L\left[\hat{\mathcal F}(\lambda_L)-\hat{\mathcal F}(\lambda_0)\right]}{\lambda_L-\lambda_0}\,,
\end{equation}
where $\hat{\mathcal F}(\lambda)=\int \hat f(\lambda)d\lambda$.

Since the external field $E_{eq}=E_{eq}(\lambda_0,\lambda_L)$ is determined by the boundary values of the chemical potentials, in this case the differential \eqref{diffF} becomes
\begin{equation}
d  F_{eq}=\frac{\partial  F_{eq}}{\partial \lambda_0}d\lambda_0+\frac{\partial  F_{eq}}{\partial \lambda_L}d\lambda_L
\end{equation}
and we have
\begin{equation}
\left\{
\begin{array}{l}
\frac{\partial  F_{eq}}{\partial \lambda_0}=\frac{L\left[\hat{\mathcal F}(\lambda_L)-\hat{\mathcal F}(\lambda_0)-\hat f(\lambda_0)(\lambda_L-\lambda_0)\right]}{(\lambda_L-\lambda_0)^2}\,,\\
\frac{\partial  F_{eq}}{\partial \lambda_L}=\frac{L\left[\hat{\mathcal F}(\lambda_0)-\hat{\mathcal F}(\lambda_L)-\hat f(\lambda_L)(\lambda_0-\lambda_L)\right]}{(\lambda_0-\lambda_L)^2}\,.
\end{array}
\right.
\end{equation}
In the case of independent particles we have $f(x)=x\log x-x$ and by a direct computation we obtain $\hat{\mathcal F}(x)=e^x(x-2)$ while in the case of exclusion process we have $f(x)=x\log x +(1-x)\log(1-x)$ and again by a direct computation we have $\hat{\mathcal F}(x)=\operatorname{Li}_2(-e^x)+\log(e^x+1)$; where $\operatorname{Li}_2(z)$ is the polylogarithm of order $2$ defined by $\operatorname{Li}_s(y):=\sum_{k=1}^{+\infty}\frac{y^k}{k^s}$.

\medskip

We can also consider equilibrium states on which the external field is not constant and in dimension greater than one. This situation corresponds to the case when the external field $E(x)=\nabla U(x)$ with $U$ a function such that $U(x)=\lambda(x)$ for $x\in \partial \Lambda$. In this case we have also that
$\bar \rho(x)= \hat f (U(x))$.

To have an equilibrium state of this type in the pair $(E,\lambda)$ the external field and the chemical potential must be related one with the other and the equilibrium state can be parameterized naturally by the function $U$ considering $F_{eq}(E,\lambda)=F_{eq}(U)$. We have
\begin{equation}
F_{eq}(U)=\int_\Lambda  \hat{f} (U(x))\,dx\,,
\end{equation}
and therefore we get
\begin{equation}
dF_{eq}=\int_{\Lambda}\frac{\delta F_{eq}(U)}{\delta U(x)}\delta U(x) dx=\int_{\Lambda} \hat{f}'(U(x))\delta U(x) dx\,.
\end{equation}

\subsection{Independent particles}
In the case of independent particles we have that $D(\rho)=1$, $\chi(\rho)=\rho$ and $f(\rho)=\rho\log\rho-\rho$ and we discuss the general case considering also non equilibrium situations. We can then compute
\begin{equation}
G_{c,E}(\rho)=\frac 1E \log \frac{|E\rho -c|}{|E\rho_0 -c|}\,,
\end{equation}
from which we get that the constant c is given by
\begin{equation}
\hat J(\lambda,\rho)=c=\frac{E(e^{\lambda_L}-e^{\lambda_0+EL})}{1-e^{EL}}\,.
\end{equation}
From formula \eqref{staz-sol} we obtain
\begin{equation}
\bar\rho(x)=e^{\lambda_0+Ex}+\frac{e^{\lambda_L}-e^{\lambda_0+EL}}{1-e^{EL}}\left(1-e^{Ex}\right)\,,
\end{equation}
that can be written as
\begin{equation}\label{ab}
\left\{
\begin{array}{l}
\bar\rho(x)=ae^{Ex}+b\\
a=\frac{e^{\lambda_0}-e^{\lambda_L}}{1-e^{EL}}\,,\\
b=\frac{e^{\lambda_L}-e^{\lambda_0+EL}}{1-e^{EL}}\,.
\end{array}
\right.
\end{equation}
We deduce therefore, with a change of variables, the form of the free energy
\begin{equation}\label{free-free}
F(E,\lambda)=\int_1^{e^{EL}}dy \frac{(ay+b)(\log(ay+b)-1)}{Ey}\,.
\end{equation}
A more explicit form can be obtained using the fact that the primitive of the function appearing in the integral \eqref{free-free} is
\begin{equation}
\mathcal G_{a,b}(y):=\dfrac{\ln\left(ay+b\right)\left(b\ln\left(\frac{\left|a\right|\left|y\right|}{\left|b\right|}\right)+ay+b\right)-b\ln\left(\left|y\right|\right)+b\operatorname{Li}_2\left(\frac{ay+b}{b}\right)-2ay}{E}\,.
\end{equation}
 We have then
\begin{equation}
 F(E,\lambda)=\mathcal G_{a,b}(e^{EL})-\mathcal G_{a,b}(1)\,,
\end{equation}
where the constant $a,b$ are defined in \eqref{ab}.



\subsubsection{Formula \eqref{f} for $D=1$ and $E=0$}

In this case we perform the computations solving equations \eqref{davide2bis}.
The Einstein relation gives $f''=\chi^{-1}$, and the stationary solution is linear $\bar \rho_s(x)=\rho_0(s)+\frac{x}{L}\left(\rho_L(s)-\rho_0(s)\right)$. The equation for $r$ is given by
\begin{equation}
\left\{
\begin{array}{l}
\partial_s\bar\rho_s=\partial^2_x r\\
r(s,0)=r(s,L)=0
\end{array}
\right.
\end{equation}
Since the left hand side of the first equation above is a first order polynomial in $x$ then the solution is a third order polynomial in $x$
\begin{equation}
r(s,x)= R_3(s)x^3+R_2(s)x^2+R_1(s)x+R_0(s)
\end{equation}
and by a direct inspection of the constraints we get
\begin{equation}
\left\{
\begin{array}{l}
R_3(s)=\frac{\dot \rho_L(s)-\dot\rho_0(s)}{6L}\\
R_2(s)=\frac{\dot\rho_0(s)}{2}\\
R_1(s)=-L\frac{\dot\rho_L(s)+2\dot\rho_0(s)}{6}\\
R_0(s)=0\,,
\end{array}
\right.
\end{equation}

By using the arguments in section \eqref{sec:ex} we can obtain the form of the free energy in this case that is
\begin{equation}\label{FzeroE}
F(\rho)=F(\rho_0,\rho_L)=\int_0^L f\left(\mathcal D^{-1}\left(\mathcal D(\rho_0)+\frac{\mathcal D(\rho_L)-\mathcal D(\rho_0)}{L}x\right)\right)dx\,,
\end{equation}
where $\mathcal D(\alpha):=\int_0^\alpha D(y)dy$. In the case of a model with a constant diffusion, like independent particles or the exclusion process, formula \eqref{FzeroE} becomes
\begin{equation}\label{x}
F(\rho)=F(\rho_0,\rho_L)=\int_0^L f\left(\rho_0+\frac{\rho_L-\rho_0}{L}x\right)dx\,,
\end{equation}
with $f(\alpha)=\alpha\log\alpha$ for independent particles and $f(\alpha)=\alpha\log\alpha+(1-\alpha)\log(1-\alpha)$ for the exclusion process.

\smallskip
In this case the identity \eqref{f} for quasistatic transformations can be verified with a direct computation of all the terms.

\section{Concluding remarks}

In this paper we discussed the generalization of Clausius statement on the equivalence of quasi static transformations between two fixed initial and final equilibrium states to the stationary states of a class of dissipative systems, the diffusive systems, using a formula previously derived in \cite{mft,qc}. In this way the difference of equilibrium free energy $F$ becomes an invariant under variations of the protocols in quasistatic transformations. We give here a direct proof.

\smallskip

Equivalent transformations can be distinguished in terms of the $1/\tau$ correction to the renormalized work introduced in \cite{45,46,mft,qc},  which confirms the reasonableness of our definitions.

\smallskip

It may be interesting to investigate if there is any connection with recent works, \cite{M,MS,SY,SSY} and references therein, who consider
entropy as a Noether invariant.

\section*{Declarations}
The authors have no conflicts of interest. Data sharing is not applicable to this article as no datasets were generated or analysed during the current study.

\subsection*{Acknowledgements}
We wish to acknowledge our long standing collaboration with L. Bertini, A.De Sole, C. Landim. DG acknowledges financial support from the Italian Research Funding Agency (MIUR) through
PRIN project ``Emergence of condensation-like phenomena in interacting particle systems: kinetic and lattice models'', grant n. 202277WX43.

\end{document}